\documentclass[conference]{IEEEtran}
\IEEEoverridecommandlockouts
\usepackage{cite}
\usepackage{amsmath,amssymb,amsfonts}
\usepackage{algorithm,algpseudocode}

\usepackage{graphicx}
\usepackage{textcomp}
\usepackage{xcolor}
\usepackage{color}
\usepackage{xspace}
\usepackage{url}
\usepackage{listings}
\usepackage{amsmath}
\usepackage{tabularx}
\usepackage{booktabs}
\usepackage{multirow}
\usepackage{pifont}
\usepackage{subcaption}
\usepackage{multicol}
\def\BibTeX{{\rm B\kern-.05em{\sc i\kern-.025em b}\kern-.08em
    T\kern-.1667em\lower.7ex\hbox{E}\kern-.125emX}}

\begin{document}

\newcommand{\approachname}{\emph{ASSS}\xspace}
\newcommand{\vtblptr}{\emph{vtblptr}\xspace}
\newcommand{\thisptr}{\emph{thisptr}\xspace}
\newcommand{\vcall}{\emph{vcall}\xspace}

\newcommand{\ie}{i.\,e.}
\newcommand{\eg}{e.\,g.}
\newcommand{\andre}[1]{\textit{{\color{red} Andre: #1}}}
\newcommand{\thorsten}[1]{\textit{{\color{green} Thorsten: #1}}}
\newcommand{\erik}[1]{\textit{{\color{blue} Erik: #1}}}

\newcommand{\txt}[1]{\texttt{#1}}
\newcommand{\tabh}[1]{{\bfseries #1}}
\newcommand{\tabsh}[1]{{\itshape #1}}
\newcommand{\tabshc}[1]{\#{\itshape \hspace{1pt}#1}}
\newcommand{\tabshp}[1]{+{\itshape \hspace{1pt}#1}}

\newcommand{\specialcell}[2][c]{\begin{tabular}[#1]{@{}c@{}}#2\end{tabular}}

\def\first{(i)\xspace}
\def\second{(ii)\xspace}
\def\third{(iii)\xspace}
\def\fourth{(iv)\xspace}
\def\fifth{(v)\xspace}
\def\sixth{(vi)\xspace}

\definecolor{medblue}{rgb}{0,0,0.5}
\definecolor[named]{NewtonRed1}{cmyk}{0,0.90,0.86,0}
\definecolor[named]{NewtonRed2}{cmyk}{0,0.84,0.76,0.40}
\definecolor{lightergray}{gray}{0.85}
\definecolor[named]{ACMPurple}{cmyk}{0.55,1,0,0.15}
\definecolor{dkgreen}{rgb}{0,0.6,0}
\definecolor{gray}{rgb}{0.5,0.5,0.5}
\definecolor{mauve}{rgb}{0.58,0,0.82}
\definecolor[named]{ACMDarkBlue}{cmyk}{1,0.58,0,0.21}
\definecolor{myred}{rgb}{0.545098,0.10196,0.0549}
\definecolor{gray75}{gray}{0.75}

\lstset{
	frame=none,
	language={},
	numbers=none,
	basicstyle={\small\ttfamily},
	showstringspaces=false,
	columns=flexible,
	upquote=true,
	breaklines=true,
	breakatwhitespace=true,
	tabsize=2,
	keywordstyle=\color{NewtonRed1}\bfseries,
	keywordstyle={[2]\color{NewtonRed2}\bfseries},
	commentstyle=\color{ACMPurple},
	numberstyle=\tiny\color{gray},
	belowskip=-1mm,
	escapeinside={(*@}{@*)}
}

\lstdefinelanguage{newton}{
	morekeywords={for,template,typename},
	morekeywords={[2]double,unsigned,int,inline,void,char,const,size_t},
	sensitive=false,
	morecomment=[l]{//},
	morecomment=[s]{/*}{*/},
	morestring=[b][\color{medblue}]",
	alsoletter=-,
}

\title{Towards Automated Application-Specific \\Software Stacks\\
}

\author{
	\IEEEauthorblockN{
		Nicolai Davidsson\IEEEauthorrefmark{1},
		Andre Pawlowski\IEEEauthorrefmark{2},
		Thorsten Holz\IEEEauthorrefmark{2}\\[2ex]
	}
	\IEEEauthorblockA{
		\IEEEauthorrefmark{1}
		Google, Switzerland, ndavidsson@google.com
	}
	\IEEEauthorblockA{
		\IEEEauthorrefmark{2}
		Ruhr-Universit\"at Bochum, Germany, \{andre.pawlowski,\,thorsten.holz\}@rub.de
	}
}

\maketitle

\begin{abstract}

Software complexity has increased over the years. One common way to tackle this
complexity during development is to encapsulate features into a shared library.
This allows developers to reuse already implemented features instead of
reimplementing them over and over again.
However, not all features provided by a shared library are actually used by an application.
As a result, an application using shared libraries loads unused code into memory,
which an attacker can use to perform code-reuse and similar types of attacks.
The same holds for applications written in a scripting language such as PHP or Ruby:
The interpreter typically offers much more functionality than is actually required by the application and
hence provides a larger overall attack surface.

In this paper, we tackle this problem and propose a first step towards automated application-specific software
stacks. We present a compiler extension capable of removing unneeded code from shared libraries
and---with the help of domain knowledge---also capable of removing unused functionalities from an interpreter's code base
during the compilation process.
Our evaluation against a diverse set of real-world applications, among others \emph{Nginx},
\emph{Lighttpd}, and the PHP interpreter, removes on average 71.3\% of the code in
\emph{musl-libc}, a popular libc implementation.
The evaluation on web applications show that a tailored PHP interpreter can mitigate
entire vulnerability classes, as is the case for \emph{OpenConf}.
We demonstrate the applicability of our debloating approach by creating an application-specific software stack for
a Wordpress web application: we tailor the libc library to the Nginx web server and PHP interpreter, whereas
the PHP interpreter is tailored to the Wordpress web application. In this real-world scenario,
the code of the libc is decreased by 65.1\% in total, thereby reducing the available code for code-reuse attacks.

\end{abstract}


\section{Introduction}


To reduce complexity of software and provide
low-level features in a consistent manner, the concept of shared libraries was developed.
This gives developers the possibility to focus solely on the user-facing
application rather than re-implementing common functionality such as memory management or string processing functions over and over again.
However, since not all code of a given shared library is used in a given program,
the downside of this concept is that unnecessary code
is loaded into memory: a recent study finds that only
5\% of the \emph{libc}, the standard library for the C programming language, is used on average across 2,016 applications of the
Ubuntu Desktop environment~\cite{quach2018debloating}.

From an attacker's perspective, the typical way to exploit an existing vulnerability is
to reuse existing code (\eg, ret2libc~\cite{tran2011expressiveness}
or return-oriented programming~\cite{shacham2007geometry} (ROP)) to execute
shellcode and bypass existing mitigation systems such as W$\oplus$R
and address space layout randomization (ASLR).
Since shared libraries offer a plethora of (mostly) unused code,
the attacker has a large variety of existing functions or code parts to choose from.

The same holds for applications written in interpreted languages, such as PHP, Python, or Ruby:
the interpreter is a complex piece of software and offers more functionality than the application actually requires~\cite{quach2017multi}.
Hence, an attacker that is able to inject her own script code into the application
can leverage these provided but unused methods to execute her exploit.


One way to remove the unused code of a shared library is to statically link it
against the target application. This allows the linker to remove the unnecessary
code and thus reduce the availability of code snippets an attacker can choose from for a code-reuse attack. However, this increases the complexity
in managing software updates: since each application has to be compiled statically linked
with all used libraries, each has to be updated when a vulnerability is found
in the code of \emph{any} used library.
To tackle this problem, Quach~et~al.~\cite{quach2018debloating} presented the concept of \emph{piece-wise compilation
and loading}. It allows to compile an application and shared libraries with additional metadata
to have a customized loader only load the needed code into memory.
Unfortunately, the concept of this approach only works with shared libraries and does not apply
to applications written in interpreted languages.


In this paper, we present a first step towards \emph{automatic application-specific software stacks}.
Our goal is to customize the software stack for a given application (\eg, a web application or server application)
such that only the actually required library code and underlying execution environment is contained within the software stack,
hence \emph{debloating} the software stack.
To achieve this goal, we introduce a compiler extension capable of removing
unused code from shared libraries, written in C. 
With information about which exported functions the target application uses, the compiler pass can omit
functions at compile time from the shared library that are not used by the application or library itself.
As a result, a shared library specifically tailored to the target application
is created. To enhance usability, our approach is able to create shared libraries that
are tailored to more than one application (\eg, a script interpreter and a web server).
In contrast to a statically linked library, tailoring to a group of applications
provides the same flexibility as a dynamically shared library
given that only the shared library has to be re-compiled if a vulnerability in its code was discovered.
When deployed with other existing defenses, such as Control-Flow Integrity
(CFI)~\cite{abadi2005control}, an application-specific software stack
further restricts the wiggle room an attacker can exploit to perform a successful attack.

Moreover, we show that---with the help of domain knowledge---this approach is also capable of removing unused functionalities in script interpreters
when targeting an application written in an interpreted language (such as PHP or Ruby).
Consider for example a Wordpress installation. With our approach,
a PHP interpreter can be tailored to the concrete Wordpress web application.
Since all unused functionalities are removed from the interpreter, an attacker that
is able to inject script code (\eg, by uploading a script file) is no longer able to leverage them for their attack.
Moreover, instead of removing unused functionalities in the interpreter, our approach allows to replace them with
\emph{booby traps}~\cite{crane2013booby}, \ie, dormant code that when executed triggers an alarm.
This way, an ongoing attack can be detected when a functionality that was removed is executed.
Note that the Wordpress-specific PHP interpreter and the web server can be compiled with our debloating approach for libraries,
leading to an application-specific software stack.
Regarding the recent trend to separate services into container (such as Docker~\cite{docker})
to provide a better security in case of a vulnerability, this makes tailoring shared libraries to
specific server applications real-world deployable.

An application-specific script interpreter also allows to reduce the
attack surface significantly in environments in which untrusted
scripts are executed (such as Google App Engine~\cite{googleappengine}).
Normally, unwanted functionalities are disabled in configuration files. However, since
the code that provides these functionalities is still available in the script interpreter,
an attacker might be able to bypass the restrictions and escape the interpreter's internal sandbox~\cite{park2018bytecode}.
When compiling the script interpreter in an application-specific way, the code for the
unneeded functionalities are completely removed, which prevents an attacker from using them entirely.


We evaluated our prototype compiler pass for LLVM by tailoring two libc implementations
(\emph{musl-libc} and \emph{uClibc}) to a diverse set of applications. The results show
that on average the code for the \emph{musl-libc} tailored to an application is reduced by 71.3\%.
A previous study on libc utilization~\cite{quach2018debloating} concluded that only 5\% of code
on average is used in the library. However, their evaluation set consists of mostly small applications, which explains
the significant difference in comparison to our results.
Additionally, we show that by using domain knowledge, our prototype is able to mitigate possible attacks on web applications: starting from \emph{seven} security-critical PHP functions that might be used 
for \emph{remote command execution} (according to the RIPS code analyzer~\cite{rips}) in the interpreter,
a PHP interpreter tailored to \emph{OpenConf} or \emph{FluxBB} only contains \emph{one} sensitive PHP function.
This significantly raises the bar for an attacker able to execute own PHP code since using a removed PHP functionality triggers
a booby trap and hence raises an alarm. In fact, in case of \emph{OpenConf}, our approach removes
the possibility to execute shell commands from the interpreter in most system configurations due to the nature of
the remaining sensitive PHP function.
Additionally, we show the real-world applicability of our approach by creating a Docker container consisting of
an application-specific software stack for a \emph{Wordpress} installation.
Our evaluation shows that the code of the libc used by the web server and PHP interpreter in this container
is reduced by 65.1\% in total, hence demonstrating that our debloating approach removes a signification fraction of unused code.

\smallskip \noindent\textbf{Contributions}.
In summary, we provide the following contributions in this paper: 

\begin{itemize}
	\item We present the design and implementation of an LLVM compiler pass capable of removing unused code from shared libraries and
	      script interpreters written in C that effectively reduces the available code snippets for reuse attacks by debloating the software stack used by a given application.
	\item Our evaluation shows that on average 71.3\% of the code in the \emph{musl-libc} is removed when tailoring it
	      to a target application. Moreover, when applying our approach to the PHP interpreter by targeting
	      specific web applications, it is capable of eliminating entire vulnerability classes, such as \emph{command execution}.
\end{itemize}

To foster research on this topic, we release the prototype implementation of our LLVM compiler pass 
as open-source software under \url{https://github.com/RUB-SysSec/ASSS}.

\section{Background}

Shared libraries offer developers a way to reuse already implemented functionalities in their program.
These functionalities can either be \emph{code} in the form of functions or \emph{data} (\eg, global variables).
For example, libc provides the developer with a variety
of low-level functionalities (\eg, memory allocation and string processing).
During compilation, there are two ways to couple the external functionalities with the own application:
\emph{static linking} and \emph{dynamic linking}.
In case of static linking, the external functionalities are resolved and plainly copied into the application during
compilation. This means that no shared library is needed to execute the application since all library-provided
functionalities are part of the program itself and hence available in memory.
In case of dynamic linking, the external functionalities are replaced with a symbol which is resolved during
the execution of the program. Hence, the shared libraries that provide
the functionalities have to be present in memory to execute the application.

In practice, dynamic linking is used in most deployment scenarios. This allows the system to use the same shared library
for multiple applications. Furthermore, having only one copy of the shared library improves usability during software patching:
if a vulnerability is found in a function offered by a shared library, the user only needs to update the corresponding shared library.
Since all dependent applications use this shared library, the vulnerability is fixed for all of them. In case of
static linking, all applications using this functionality have to be updated to fix the vulnerability.
As explained earlier, the main downside of using dynamic linking is the fact that this approach increases the amount of 
unused code that is mapped into the memory of the application.
Therefore, sensible operations in functionalities not used by the application itself are also present in memory.

\section{High-Level Overview}

\begin{figure*}[t]
	\centering
	\includegraphics[width=.99\textwidth]{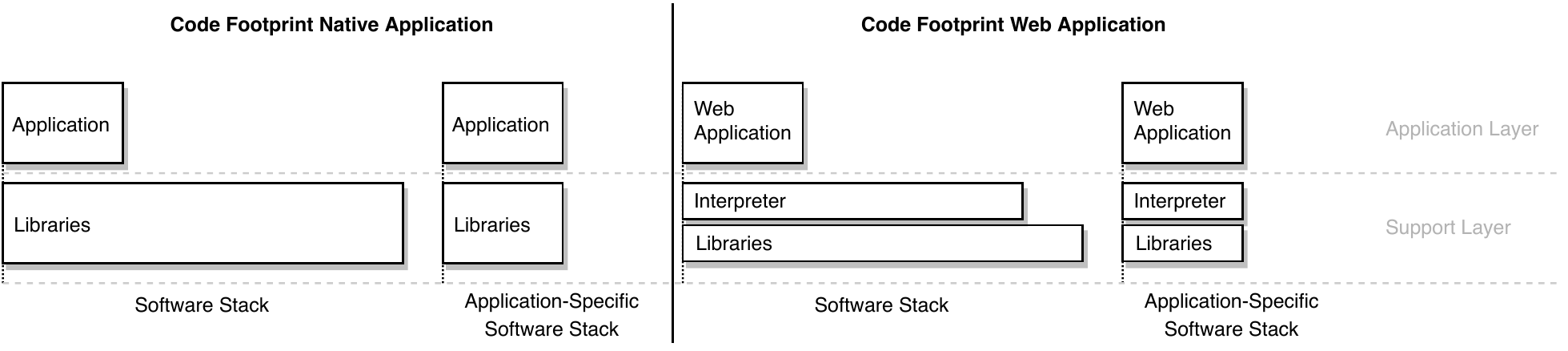}
	\caption{Schematic overview of the difference of the code footprint between a traditional software stack
		and an application-specific one showing the average code reduction of around 70\%.
		On the left it is shown for a native code application, on the right for an interpreted
		application such as a web application. Even though the interpreter is part of the support layer, it also
		does not use the complete code provided by the shared libraries.}
	\label{fig:overview}
\end{figure*}

The idea behind application-specific software stacks is based on the observation that applications
do not use every functionality provided by their underlying software stack (\eg, interpreters or libraries).
Therefore, it is safe to remove code of these unused functionalities
to debloat the application without affecting it. Furthermore, by removing code snippets or whole functions that can potentially be used by an attacker
in code-reuse attacks narrows down the options an attacker has.
This also holds for scripting languages, for example, in a web application
context: the script interpreter offers more functionality than
the web application uses. Stripping the interpreter from these functionalities
debloats the interpreter, but does not interfere with the given web application.
Moreover, in cases an attacker is able to insert her own script code (\eg, by uploading a script file to a web server),
she is limited in the interpreter functionalities she can use. Figure~\ref{fig:overview} shows the difference between a
normal software stack and an application-specific one graphically.

We define two layers for a software stack: the \emph{application layer} and the \emph{support layer}.
The application itself resides on the application layer. This can either be
a native code application or an application written in an interpreted language (\eg, web application).
In a web application context, the application layer also includes the web framework the application uses.
The libraries and script interpreter are located on the support layer. 
This layer provides functionalities that are used by the application.
However, it also contains additional code and functionalities that are not used by the application.
Underneath the support layer resides the operating system (OS).
Functionalities provided by the OS are usually accessed via the support layer through low-level libraries such as the libc.

Our goal is to debloat the software stack by removing unneeded code from the support layer.
This is done by analyzing the application and retrieving
control transfers from the application layer into the support layer.
This information is then used to recompile the support layer without
the unused code. The result is a software stack tailored
to the application. However, this approach is not limited to tailoring the
support layer to only one application, thus increasing its usability.
Consider for example a Wordpress installation. The libraries used by the web server and PHP interpreter
can be specifically tailored to support both. Moreover, the PHP interpreter can be customized to only
contain functionalities used by Wordpress. Hence, the debloating is achieved throughout the
whole software stack by preserving the usability of shared libraries.

In the case of native code applications, the same code reduction can be achieved by using
static linking during the compilation and linking process. As a result, the functionalities provided by the libraries
and used by the application are directly inserted into the code of the program. This moves part of the support layer
directly into the application layer. However, this also means that the advantages of sharing libraries between
multiple applications are also lost. As a result, as soon as a vulnerability is discovered in a library
functionality, all applications using this library have to be updated. 
Application-specific software stacks, on the other hand, still provide the advantages of shared libraries.
It is possible to group different applications to use one shared library tailored to them (as in our example
a web server and script interpreter).
Hence, our approach offers a middle ground between code reduction and usability.

\section{Approach}

In this section, we describe our approach for application-specific software stacks.
We start by describing the basic method of our LLVM pass and refining it step-by-step
throughout this section until each challenge encountered is tackled.
The final goal in this paper is to create a Wordpress installation with
a tailored PHP interpreter and a libc implementation application-specific
to the interpreter and web server.
Hence, the described method focuses only on tailoring the libraries to a target
application first. Afterwards, domain knowledge is used to enhance our approach
to also support specific script interpreters, more specifically, the PHP and Ruby interpreter.
The complete algorithm is shown in Algorithm~\ref{alg:complete} in the Appendix.

\subsection{Libraries}
\label{sec:nativecode}

The control-flow transfer from the application layer into the support layer
can be performed in multiple ways. In the easiest form, it is a
direct call of a function. However, more complicated constructs such
as indirect calls via function pointers are also possible.
An analysis tailoring libraries to a specific application at compile time 
must not miss any of these, since one missing functionality leads to
an uncompilable library in the best case, and a broken application in the worst.
Next, we describe a method for LLVM capable of handling all these cases.

\paragraph{Base Method}

\begin{figure}
	\centering
	\includegraphics[width=.80\columnwidth]{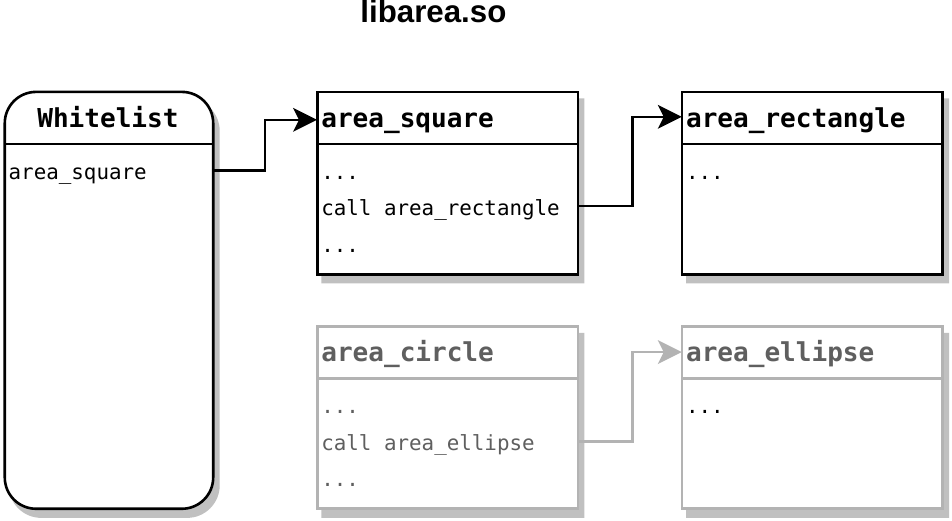}
	\caption{Example of the basic idea of the analysis. A target application uses the function \texttt{area\_square}. Hence,
		the function \texttt{area\_rectangle} is also added to the whitelist.}
	\label{fig:base_algo}
\end{figure}

We start with a whitelist of functions, which initially
contains all exported functions of the library used by the target application.
The exported functions can be obtained by reading the metadata of the target application (\eg, with the help of the
binutils tool \texttt{readelf}).
Consider the example shown in Figure~\ref{fig:base_algo}. The target application uses the function \texttt{area\_square}
of the library. During compilation, each currently processed function is checked if it resides in the whitelist.
If the function \texttt{area\_square} is processed, all direct control-flow transfers are also explored.
Each new function that is reachable by the direct control-flow transfer is added to the whitelist
and further explored. In this example, the function \texttt{area\_rectangle} is added to the whitelist.
This phase of searching for new reachable
functions is called \emph{function exploration}. Since this phase uses a
depth-first search (DFS) approach,
it is guaranteed to visit all functions that are reachable by an initial given function.
Hence, all functions in the whitelist after the analysis is finished
are necessary for the application to work. All other functions can be safely removed.
In the given example, \texttt{area\_circle} and \texttt{area\_ellipse} are dismissed.

\paragraph{Indirect Control-Flow Transfers}
Unfortunately, the compiler cannot always determine the target of a control-flow
transfer. Often control-flow transfers are handled with the help of function pointers,
\ie, through indirect call instructions. Therefore, 
we have to consider them during our analysis.
Hence, we have to extend our approach to work with instructions handling function pointers. 
We found that the following LLVM intermediate representation (IR) instructions are
capable of handling function pointers:
\begin{itemize}
	\item \texttt{store}: storing data in a variable.
	\item \texttt{return}: returning data at the end of a function.
	\item \texttt{select}: chooses between two distinct values depending on a boolean condition.
	\item \texttt{phi}: merging multiple variables into a single variable for Single Static Assignment (SSA) form~\cite{cytron1991efficiently}.
\end{itemize}
Since all these instructions can work with a function pointer,
our analysis has to be able to process them.
Therefore, we extend the \emph{function exploration} phase to
extract the data handled by these instructions to find
all indirect control-flow transfers. If the extracted data is a function pointer,
we continue the exploration at the pointer target.

This refined method handles all possible function pointers that are
set inside the used code. However, since the semantics of the code
are not considered, this analysis can overestimate the actually
used functions. Consider for example a \texttt{select} instruction
that chooses between two function pointers. When the boolean
condition evaluates always to true, then only one function
is ever reached by this code construct. Yet, our analysis
considers both functions as reachable and thus overestimates
the actually used functions. 
Note that this conservative overestimation guarantees us to
not break the application.

\paragraph{Global Variables}

\begin{figure}
	\small
	\centering
	
	\begin{subfigure}{\columnwidth}
		\begin{lstlisting}[
		language=newton,
		%xleftmargin=2.0em, % otherwise line numbers are in text
		frame=single,
		numbers=none,
		basicstyle=\ttfamily\small,
		firstnumber=7,
		breaklines=true]
		static FILE f = {
		    .write = __stdout_write,
		    //...
		};
		FILE *const stdout = &f;
		\end{lstlisting}
		\vspace{2mm}
		\caption{Definition of a function pointer in a global variable from \texttt{src/stdio/stdout.c}.}
		\label{lst:filestruct}
	\end{subfigure}
	
	\vspace{3mm}
	
	\begin{subfigure}{\columnwidth}
		\begin{lstlisting}[
		language=newton,
		%xleftmargin=2.0em, % otherwise line numbers are in text		
		frame=single,
		numbers=none,
		basicstyle=\ttfamily\small,
		firstnumber=7,
		breaklines=true]
		size_t __fwritex(const unsigned char *restrict s, size_t l, FILE *restrict f)
		{
		    //...
		    size_t n = f->write(f, s, i);
		    //...
		}
		
		size_t fwrite(const void *restrict src, size_t size, size_t nmemb, FILE *restrict f)
		{
		    //...
		    k = __fwritex(src, l, f);
		    //...
		}
		\end{lstlisting}
		\vspace{2mm}
		\caption{Possible usage of a function pointer from \texttt{src/stdio/fwrite.c}.}
		\label{lst:fwrite}
	\end{subfigure}
	\caption{Code snippets from \emph{musl-libc} that show the usage of function pointers in global variables.}
\end{figure}

Although function pointers set directly in the code are already handled by
our analysis, function pointers can also reside in global variables.
Consider the code snippet from \emph{musl-libc} shown in Figure~\ref{lst:filestruct}.
The function \texttt{\_\_stdout\_write} is stored as function pointer
in the global variable \texttt{stdout}. This variable is a \texttt{FILE} struct
used for I/O operations.
Figure~\ref{lst:fwrite} shows the function \texttt{fwrite}
that uses a pointer to a \texttt{FILE} struct to invoke the \emph{write}
function stored there.
Since the current form of our algorithm is not able to find the 
\texttt{\_\_stdout\_write} function pointer in the global variable,
a valid call to \texttt{fwrite} with the global variable 
\texttt{stdout} as argument would break the application.

To handle global variables, we add a \emph{global exploration} phase
to our analysis. In this phase, all global variables are processed and checked
for function pointers. If they contain a function pointer, the target is added
to the whitelist as well. 
The \emph{global exploration} phase is executed before the \emph{function exploration} phase
to guarantee that the newly whitelisted functions are also explored.
A discussion about limitations of our function pointer analysis is given in Section~\ref{sec:discussion}.

\subsection{Script Interpreters}

Often applications written in scripting languages like PHP, Ruby, or Python
are not translated
into native code, but interpreted by the corresponding script interpreter.
As a result, the interpreter itself is a part of the support layer
for these applications. However, in contrast to the method described
in Section~\ref{sec:nativecode} for native code libraries, the analysis
cannot just remove code from the interpreter since it cannot distinguish
which code belongs to a certain interpreter functionality.
Hence, to build an application-specific interpreter, our analysis
has to leverage domain knowledge about the internals of the target interpreter.
More specifically, the analysis has to know the mapping of script functions to native code functions.
To achieve our goal of running a Wordpress installation with an application-specific
interpreter, we modify our analysis to work with the PHP interpreter in the following.
To show that our approach is
not limited to PHP, we further extend our algorithm to work with the Ruby
interpreter.

\paragraph{PHP}
\label{sec:approachphp}
PHP stores information for each registered PHP function in global \emph{function entries},
which are basically a map of structs~\cite{regphpfuncs}.
The structs contain, among others, the pointer to the native code function and the 
name of the PHP function. During execution, they are used to handle the
transition from PHP to native code. The interpreter uses these function entries to
look up the native code function that is eventually executed to perform the
application's desired functionality.
Hence, modifying these function entries during the compilation of the PHP interpreter
to remove the code from it is the best way to keep our approach as generic as possible.
Since the function entries are part of the architecture of PHP,
they are less likely to change between different PHP versions and hence
our approach should be compatible with upcoming PHP releases.

To enable our analysis to remove PHP functionalities from the interpreter
at compile time, we introduce a whitelist of PHP functions and
modify the \emph{global exploration} phase. The modification extracts
the PHP function names from the PHP global \emph{function entries} and checks if they
are on the PHP whitelist. If they are, the corresponding native function is stored
for processing during the \emph{function exploration} phase.
As a result, the native code corresponding to the functionality
only remains in the interpreter when it is on the PHP whitelist.

PHP supports the paradigm of object-oriented programming, \ie, 
functions can be associated to classes. An example of a class
and its member function directly provided by the PHP interpreter
is the \texttt{Directory} class and its function
\texttt{read}~\cite{phpdirread}.
However, the PHP function name
does not contain any information about the associated class. Hence,
if multiple classes register a PHP function with the same name,
our analysis is not able to distinguish between them.
Consider an example where classes \texttt{A} and \texttt{B} both register
a function with the name \texttt{read}.
If the application only uses \texttt{A::read}, our analysis will
still whitelist the \texttt{read} function of both classes.
This loss in precision results in the PHP interpreter still containing
functionality that is not needed, however, it guarantees to not
break the application.

\paragraph{Ruby}

\begin{figure}
	\begin{lstlisting}[
	language=newton,
	%xleftmargin=2.0em, % otherwise line numbers are in text
	frame=single,
	numbers=none,
	basicstyle=\ttfamily\small,
	firstnumber=7,
	breaklines=true]
	void Init_IO(void)
	{
        //...
        rb_define_global_function("syscall",
                                           rb_f_syscall,
                                           -1);
        rb_define_global_function("open",
                                           rb_f_open,
                                           -1);
        rb_define_global_function("printf",
                                           rb_f_printf,
                                           -1);
        //...
	}
	\end{lstlisting}
	\caption{Ruby function registration example from \texttt{io.c}.}
	\label{lst:rubyregister}
\end{figure}

In contrast to PHP, Ruby does not register its functions through global
function tables. Instead, Ruby functions are added by calling internal
register functions at runtime. Figure~\ref{lst:rubyregister} 
shows how various I/O functions are registered in the Ruby source code.
The arguments to the internal register
function contain the name of the Ruby function and a pointer to its native code pendant.
A list of all internal register functions can be found in Listing~\ref{lst:rubyregisterfcs}.
To remove the functionality provided by Ruby functions,
we aim to remove these registering function calls from the code.
Again, our approach tries to focus on the architecture of the
interpreter since this is less likely to change between versions
and hence our approach should be compatible with upcoming Ruby releases.

\begin{minipage}{0.9\linewidth}
	\begin{lstlisting}[
	language=C,
	caption={A list of Ruby's internal registration functions.\vspace{0.4em}},
	xleftmargin=2.0em, % otherwise line numbers are in text
	label=lst:rubyregisterfcs,
	frame=single,
	numbers=left,
	basicstyle=\small,
	firstnumber=1,
	breaklines=true,
	language=C,
	basicstyle=\ttfamily\small]
	rb_define_protected_method
	rb_define_private_method
	rb_define_singleton_method
	rb_define_method
	rb_define_method_id
	rb_define_module_function
	rb_define_global_function
	rb_define_alloc_func
	rb_define_virtual_variable
	rb_define_hooked_variable
	\end{lstlisting}
	\vspace{3.5em}
\end{minipage}

To enable our analysis to remove Ruby functionalities from the interpreter at compile time,
we added a whitelist for Ruby functions and modified the \emph{function exploration} phase.
Basically, the \emph{function exploration} phase now checks if a call instruction calls an internal register function
to register a Ruby functionality. If it does and the Ruby function name of the registration
is whitelisted, the corresponding native code function is further explored.
Otherwise, the corresponding call instruction is deleted from the code. 
As a result, only the code corresponding to whitelisted
Ruby functions is part of the compiled Ruby interpreter and since the call instruction
is removed, the Ruby interpreter does not register the functionality
at runtime and preserves its integrity.

This method works with Ruby functions registered directly in the code. However, dynamically
registered Ruby functions are not detected.
But since we did not encounter any dynamically registered Ruby functions in the core functionalities,
we did not pursue it further.

\section{Implementation}

Our prototype implementation resides inside the compiler itself since it
has to be able to modify the code and data structures directly (\eg, for the PHP interpreter).
Hence, to build a tailored software stack for a given software,
the whole support layer has to be re-compiled using our compiler pass.
The support layer consists of the libraries (and script interpreter) of the target application,
and all libraries used by the libraries.
Eventually, an application-specific software stack is created for the given software.
For native code applications, the used exported functions have to be extracted
as initial information for the compiler pass (\eg, with the help of the
binutils tool \texttt{readelf}).
For applications using script languages, the
analysis to get all used interpreter functionalities has to be done by
external tools like Parse~\cite{parse} for PHP.

We built the prototype of our approach as compiler pass for LLVM 5.0.1.
In total, our implementation consists of around 1,000 lines of C++ and 100 lines of Python code.
To prevent possible dependency issues, each created module by LLVM
is merged into one. This gives our compiler pass a global view of all
existing code and data. Since our pass works on the LLVM IR, it
is completely architecture and platform independent. Hence, each architecture
that is supported by LLVM is also supported by our approach (\eg, ARM or MIPS).

In order to integrate it into the build process of an application
as seamlessly as possible, we created a compiler wrapper script.
This script is used as compiler for the application and handles all steps needed
to perform our analysis.

\subsection{Manual Configuration}
\label{sec:manualconf}

\begin{figure}
	\begin{lstlisting}[
	language={[x86masm]Assembler},
	%xleftmargin=2.0em, % otherwise line numbers are in text
	frame=single,
	numbers=none,
	basicstyle=\ttfamily\small,
	firstnumber=7,
	breaklines=true]
	sigsetjmp:
	__sigsetjmp:
	;...
	.hidden __sigsetjmp_tail
	jmp __sigsetjmp_tail
	;...
	\end{lstlisting}
	\caption{Assembler instructions initiating a control-flow transfer
		found in \texttt{src/signal/x86\_64/sigsetjmp.s}}
	\label{lst:sigsetjump}
\end{figure}

Although our approach aims to automate the process in creating an
application-specific software stack, a user might want to preserve
certain functionality in the libraries. This can have various reasons,
\eg, using the same library by multiple applications.
Hence, the user is able to modify the configuration
file for the library and add additional function names to the whitelist.
Furthermore, a library could need an additional
whitelisted function which is not referenced directly from the application.
This is the case for C entry functions (\eg, \texttt{\_start}) which are directly
called by the loader during load time.

Since LLVM does not lift assembly instructions
into its IR, control-flow transfers to functions done in
assembly are not detected by our analysis.
Figure~\ref{lst:sigsetjump} shows an example encountered while compiling \emph{musl-libc} for 
the target application.
The control-flow transfer to function \texttt{\_\_sigsetjmp\_tail}
is not detected by our analysis.
We encountered five such cases in which assembly instructions
in the code call a function not referenced in the rest of the code base
(three in \emph{musl-libc} and two in \emph{uClibc}).
Since we did not encounter any cases outside of the libc, we believe such cases
more common in libraries providing low-level functionalities such as memory management
and hence an exception.

Another case for manual configuration are functions that are resolved dynamically via loader functionalities
such as \texttt{dlsym}. Since these functions do not have a reference in the code (either a direct reference
or an indirect via a function pointer), our current prototype is not able to detect them.
However, since we only encountered one case of dynamically resolved functions during our evaluation
(\texttt{\_\_dls3} in \emph{musl-libc}), we believe this feature to be rarely used in practice.
Furthermore, this function was not resolved by loader functionalities, but by a self-implemented version
of \texttt{dlsym} inside the \emph{musl-libc}. This shows further how difficult it is to fully automate
the process of creating an application-specific software stack and the reason for allowing manual configuration.
A detailed discussion on how to address these cases in an automated way is given in Section~\ref{sec:discussion}.

\subsection{Booby Trapping Script Interpreters}
\label{sec:booby}

Most scripting languages offer ways to list all registered functions.
An attacker able to execute script commands is therefore able
to use this functionality as information leak to circumvent removed
functionality. For example, the PHP function \texttt{get\_defined\_functions}
returns all functions registered to the interpreter.
To thwart these attempts, our approach is not only able to
remove functionality from the script interpreter, but to replace
its native code implementation with a booby trap~\cite{crane2013booby}.
A booby trap contains code that when executed warns from an attack.
Since this code lies dormant in memory and is never executed by the
benign application, an execution of this code detects an altered control flow
and hence an ongoing attack.
When the native code implementation of a script function is replaced
by this code, an attacker executing interpreter functionality that is not
used by the application otherwise is detected.
Furthermore, this removes any leak regarding the information about
functions registered to the interpreter.
If the attacker does not have access to the source code of the application (\eg, a proprietary application),
this removes the possibility to circumvent booby traps.

\section{Evaluation}
\label{sec:eval}

As a target for our applications, we use Linux on the Intel x86-64 architecture
because of its popularity as a server system.
In this section, we first evaluate the effect of an application-specific software stack on the
used shared libraries, afterwards a PHP interpreter tailored to specific
web applications is measured. Subsequently, we study the code
reduction of our approach on our running example: an application-specific software stack for a Wordpress installation. 
Finally, we perform a security evaluation
of our approach on the basis of several CVEs and discuss the performance overhead.

\subsection{Libraries}
\label{sec:evallib}

To evaluate the effect of our approach on native code applications,
we compile different libc versions as an application-specific software stack.
Unfortunately, the most common implementation \emph{glibc} is written in GNU C, an extension
of the C programming language which is not supported by LLVM~\cite{muench2016taming}.
Therefore, we resort to two other popular libc implementations: \emph{musl-libc} (1.1.18)
and \emph{uClibc} (0.9.34).
The \emph{musl-libc} focuses on speed, feature-completeness, and simplicity~\cite{musllibc}.
It is used, for instance, by the Alpine Linux distribution, which
is the distribution used for official Docker containers~\cite{dockeralpine}. The \emph{uClibc} implementation targets
microcontrollers and therefore focuses mainly on size~\cite{uclibc}
(\eg, it is used by the buildroot project~\cite{buildroot}).
We compile both libc implementations without any changes by our transformation to have
a complete shared library to compare against as an upper boundary. As a lower boundary, we
compile both implementations using our approach with a minimal configuration which contains
the least amount of functions necessary in the initial whitelist to compile the library
(5 functions for \emph{musl-libc} and 12 functions for \emph{uClibc}).

To show the effect of an application-specific shared library, we compile the libc implementation
for different applications: \emph{Micro-Lisp}, \emph{Nginx} (1.13.8),
\emph{Lighttpd} (1.4.48), \emph{Busybox} (1.28), \emph{PHP} (7.3.0-dev)
for different web applications, and \emph{Miniruby} (2.6.0-dev).
To have a small basic PHP interpreter that supports all base features of our used web applications,
we enabled support for Mysqli and zlib and disabled support for XML, iconv, PEAR, and DOM.
Additionally, the PHP interpreter is also compiled in a minimal configuration (the least
amount of functions necessary to run it) and in a complete configuration to better show the impact
of an application-specific library.
The \emph{Ruby} interpreter has the option to build a smaller version of itself called \emph{Miniruby}.
This interpreter only contains the core functionalities (YARV instruction set \cite{sasada2005yarv})
of the \emph{Ruby} interpreter. Since the difference between a complete \emph{Miniruby} interpreter and a minimal
\emph{Miniruby} are smaller, it is more suited to show the impact of our approach than the full-fledged
\emph{Ruby} interpreter. For \emph{Busybox}, we had to disable the coreutil functionalities: 
\texttt{date}, \texttt{echo}, \texttt{ls}, \texttt{mknod}, \texttt{mktemp}, \texttt{nl}, \texttt{stat}, \texttt{sync},
\texttt{test} and \texttt{usleep}. We were not able to compile \emph{uClibc} with LLVM when these features were activated
because of the dependency on buildroot. Hence, we had to modify the toolchain for \emph{uClibc} to work without buildroot.

\begin{table*}[t]
	\centering
	\caption{Results of the remaining code for \emph{musl-libc} and \emph{uClibc}.
		On top for each library, the table shows the number of functions and code size
		for the complete and minimal library. The minimal library shows the remaining
		code for a configuration which contains the minimal number of functions
		to compile the library. Following the same metrics for the library
		tailored to a specific application.}
	\scalebox{1.00}{
		\begin{tabular}{l r r r r | l r r r r}
			\toprule
			
			\tabh{Application} &
			\tabshc{Funcs} &
			\tabsh{\%} &
			\tabsh{Code Size} &
			\tabsh{\%} &
			
			\tabh{Application} &
			\tabshc{Funcs} &
			\tabsh{\%} &
			\tabsh{Code Size} &
			\tabsh{\%} \\

			\midrule
			
			musl-libc (complete) & 
			2,603 & 
			& 
			1,007 kB & 
			& 
			
			uClibc (complete) & 
			891 & 
			& 
			450 kB & 
			\\ 
			
			musl-libc (minimal) & 
			358 & 
			~13.8 & 
			116 kB & 
			~11.5 & 
			
			uClibc (minimal) & 
			164 & 
			~18.4 & 
			108 kB & 
			~23.9 \\ 
			
			\cmidrule(lr){1-10}
			
			\emph{Micro-lisp} & 
			366 & 
			~14.1 & 
			118 kB & 
			~11.7 & 
			
			\emph{Micro-lisp} & 
			168 & 
			~18.9 & 
			115 kB & 
			~25.5 \\ 
			
			\emph{Busybox} & 
			893 & 
			~34.3 & 
			345 kB & 
			~34.2 & 
			
			\emph{Busybox} & 
			388 & 
			~43.6 & 
			329 kB & 
			~73.2 \\ 
			
			\emph{Nginx} & 
			762 & 
			~29.3 & 
			276 kB & 
			~27.4 & 
			
			& 
			& 
			& 
			& 
			\\ 
			
			\emph{Lighttpd} & 
			745 & 
			~28.6 & 
			260 kB & 
			~25.9 & 
			
			& 
			& 
			& 
			& 
			\\ 
			
			\emph{PHP (Complete)} & 
			1,014 & 
			~39.0 & 
			390 kB & 
			~38.8 & 
			
			& 
			& 
			& 
			& 
			\\ 
			
			\emph{PHP (FluxBB)} & 
			817 & 
			~31.4 & 
			296 kB & 
			~29.4 & 
			
			& 
			& 
			& 
			& 
			\\ 
			
			\emph{PHP (OpenConf)} & 
			839 & 
			~32.2 & 
			326 kB & 
			~32.3 & 
			
			& 
			& 
			& 
			& 
			\\ 
			
			\emph{PHP (Wordpress)} & 
			874 & 
			~33.6 & 
			336 kB & 
			~33.4 & 
			
			& 
			& 
			& 
			& 
			\\ 
			
			\emph{PHP (Minimal)} & 
			768 & 
			~29.5 & 
			280 kB & 
			~27.8 & 
			
			& 
			& 
			& 
			& 
			\\ 
			
			\emph{Miniruby (Complete)} & 
			907 & 
			~34.8 & 
			325 kB & 
			~32.3 & 
			
			& 
			& 
			& 
			& 
			\\ 
			
			\emph{Miniruby (Minimal)} & 
			684 & 
			~26.3 & 
			221 kB & 
			~21.9 & 
			
			& 
			& 
			& 
			& 
			\\ 
			
			\bottomrule
		\end{tabular}
	}
	
	\label{tab:libc}
\end{table*}

\paragraph{Code Reduction}
Table~\ref{tab:libc} depicts the results of our measurements. As evident from the table,
the complete \emph{musl-libc} has 2,603 functions, whereas a minimal configuration only needs
358 functions (13.8\%) to be compilable. These configurations provide an upper and lower boundary
of the code reduction that is possible for a target application.
When tailoring the \emph{musl-libc} to a specific application,
\emph{Micro-lisp} needs the fewest functions from the library with 14.1\% remaining.
In fact, this configuration needs only eight functions more than the minimal configuration which is
necessary to compile the library.
A complete \emph{PHP} interpreter needs the most with 39.0\%.
On average, 30.3\% of the functions remain in the \emph{musl-libc} when tailored to
an application.
Since \emph{uClibc} focuses on being as small as
possible to work on microcontrollers, it does not have all features that the libc provides.
Therefore, only \emph{Busybox} and \emph{Micro-Lisp} of our evaluation set work with this library.
The complete library has 891 functions, whereas the minimal configuration
only has 164 (18.4\%). A \emph{uClibc} tailored to \emph{Micro-lisp} has 168, which are 18.9\% of all functions and only
four functions more than the minimal configuration possible. The \emph{Busybox} configuration has 43.6\% functions
remaining after its compilation. This shows that even a library focusing on being as small as possible
can be further reduced by our approach.
The code size confirms that the libraries did not only lose small wrapper-like functions, but that the
code is reduced in a proportional way to the number of functions present.

Removing PHP functionalities from the interpreter also influences the code required in the
underlying libc. A complete PHP interpreter has 39.0\% of the functions available in the \emph{musl-libc}
remaining, whereas a minimal PHP interpreter only needs 29.5\% of the functions in the library.
A PHP interpreter tailored to the \emph{Wordpress} web application, the largest web application of our evaluation set,
needs only 33.6\% of the functions of the \emph{musl-libc}.
On average, a PHP interpreter tailored to a web application needs only 32.4\% of the functions.
This shows that for software debloating
it is imperative to not only focus on the shared libraries itself,
but to take into account the actual application running when an interpreted language is used.

\paragraph{Code-Reuse Attacks}
\begin{table*}[t]
	\centering
	\caption{Results of our gadget evaluation for \emph{musl-libc} and \emph{uClibc}.
		On top for each library, the table shows the number of unique ROP gadgets
		for the complete and minimal library. The minimal library shows the remaining gadgets for a
		configuration which contains the minimal number of functions to compile the library.
		Following the same metrics for the library tailored to a specific application.}
	\scalebox{1.00}{
		\begin{tabular}{l r r r r r r r r r r}
			\toprule
			
			\tabh{Application} &
			\tabshc{unique} &
			\tabsh{\%} &
			\tabshc{JOP} &
			\tabsh{\%} &
			\tabshc{COP} &
			\tabsh{\%} &
			\tabshc{CP} &
			\tabsh{\%} &
			\tabsh{syscall} &
			\tabsh{\%} \\
			
			\midrule
			
			musl-libc (complete) &
			9,692 & 
			& 
			332 & 
			& 
			324 & 
			& 
			581 & 
			& 
			157 & 
			\\ 
			
			musl-libc (minimal) &
			1,578 & 
			~16.3 & 
			40 & 
			~12.1 & 
			106 & 
			~32.7 & 
			108 & 
			~18.6 & 
			81 & 
			~51.6 \\ 
			
			\cmidrule(lr){1-11}
			
			\emph{Micro-lisp} &
			1,581 & 
			~16.3 & 
			36 & 
			~10.8 & 
			113 & 
			~34.9 & 
			110 & 
			~18.9 & 
			81 & 
			~51.6 \\ 
			
			\emph{Busybox} &
			3,203 & 
			~33.1 & 
			152 & 
			~45.8 & 
			204 & 
			~62.7 & 
			252 & 
			~43.4 & 
			103 & 
			~65.6 \\ 
			
			\emph{Nginx} &
			3,196 & 
			~33.0 & 
			105 & 
			~31.6 & 
			166 & 
			~51.2 & 
			209 & 
			~36.0 & 
			106 & 
			~67.5 \\ 
			
			\emph{Lighttpd} &
			2,694 & 
			~27.8 & 
			97 & 
			~29.2 & 
			163 & 
			~50.3 & 
			224 & 
			~38.6 & 
			101 & 
			~64.3 \\ 
			
			\emph{PHP (Complete)} &
			4,012 & 
			~41.4 & 
			130 & 
			~39.2 & 
			235 & 
			~72.5 & 
			281 & 
			~48.4 & 
			106 & 
			~67.5 \\ 
			
			\emph{PHP (FluxBB)} &
			2,950 & 
			~30.4 & 
			99 & 
			~29.8 & 
			210 & 
			~64.8 & 
			222 & 
			~38.2 & 
			100 & 
			~63.7 \\ 
			
			\emph{PHP (OpenConf)} &
			3,387 & 
			~35.0 & 
			101 & 
			~30.4 & 
			201 & 
			~62.0 & 
			226 & 
			~38.9 & 
			97 & 
			~61.8 \\ 
			
			\emph{PHP (Wordpress)} &
			3,518 & 
			~36.3 & 
			133 & 
			~40.1 & 
			184 & 
			~56.8 & 
			223 & 
			~38.4 & 
			97 & 
			~61.8 \\ 
			
			\emph{PHP (Minimal)} &
			2,794 & 
			~28.8 & 
			85 & 
			~25.6 & 
			187 & 
			~57.7 & 
			195 & 
			~33.6 & 
			96 & 
			~61.2 \\ 
			
			\emph{Miniruby (Complete)} &
			3,533 & 
			~36.5 & 
			97 & 
			~29.2 & 
			181 & 
			~55.9 & 
			237 & 
			~40.8 & 
			112 & 
			~71.3 \\ 
			
			\emph{Miniruby (Minimal)} &
			2,578 & 
			~26.6 & 
			59 & 
			~17.8 & 
			176 & 
			~54.3 & 
			181 & 
			~31.2 & 
			104 & 
			~66.2 \\ 
			
			\midrule
			
			uClibc (complete) &
			6,101 & 
			& 
			663 & 
			& 
			285 & 
			& 
			546 & 
			& 
			733 & 
			\\ 
			
			uClibc (minimal) &
			1,736 & 
			~28.5 & 
			87 & 
			~13.1 & 
			75 & 
			~26.3 & 
			142 & 
			~26.0 & 
			150 & 
			~20.5 \\ 
			
			\cmidrule(lr){1-11}
			
			\emph{Micro-lisp} &
			1,724 & 
			~28.3 & 
			82 & 
			~12.4 & 
			77 & 
			~27.0 & 
			146 & 
			~26.7 & 
			150 & 
			~20.5 \\ 
			
			\emph{Busybox} &
			3,896 & 
			~63.9 & 
			315 & 
			~47.5 & 
			129 & 
			~45.3 & 
			312 & 
			~57.1 & 
			325 & 
			~44.3 \\ 
			
			\bottomrule
		\end{tabular}
	}
	
	\label{tab:roplibc}
\end{table*}

A modern way for an attacker to exploit a vulnerability in an application is to reuse existing code.
One way for an attacker is to transfer the control flow to an existing function in a library with
crafted arguments and therefore execute the behavior the attacker desires
(\eg, ret2libc attack~\cite{tran2011expressiveness}).
However, since the number of existing functions in the library is significantly reduced,
an attacker may not be able to find a function that executes the behavior she needs.
For example, in all configurations listed in Table~\ref{tab:libc}, except for \emph{Busybox} for \emph{uClibc},
the function \texttt{system} which is usually used to execute shell commands in an exploit is removed from the code.

Another way to reuse existing code for an attack is called return-oriented programming (ROP)~\cite{shacham2007geometry}.
For this exploiting technique, small code snippets called \emph{gadgets} are combined
by the attacker to build the shellcode.
Since an attacker needs a variety of different ROP gadgets to obtain the shellcode she needs,
we measured the reduction of gadgets in the library with the tool
\emph{ROPgadget}~\cite{ropgadget} in version 5.6.
While a tailored software stack alone does not prevent code-reuse attacks, this metric
gives an estimate on the limitation an application-specific
software stack imposes on ROP attacks.
Besides measuring the number of unique ROP gadgets remaining, we also measured security-sensitive
gadgets such as jump-oriented programming (JOP)~\cite{bletsch2011jump}, call-oriented
programming (COP)~\cite{carlini2014rop}, call-preceding gadgets (CP)~\cite{carlini2014rop}, and syscall gadgets~\cite{shacham2007geometry}. 

A minimal configuration of \emph{musl-libc} and \emph{uClibc} has only
16.3\% and 28.5\% of the unique ROP gadgets the complete library has. 
A tailored \emph{musl-libc} has in the worst case 41.4\% of unique ROP gadgets remaining for the
complete PHP interpreter and in the best case 16.3\% for \emph{Micro-lisp}.
For a tailored \emph{uClibc}, 28.3\% of the unique ROP gadgets remain for \emph{Micro-lisp} and
63.9\% for \emph{Busybox}. Since \emph{uClibc} is already optimized in regard to code size,
the gadget reduction was to be expected less than the one for \emph{musl-libc}.
A full overview of all remaining gadgets is given in Table~\ref{tab:roplibc}.

\medskip Overall, our evaluation shows that an application-specific library loses most of its code.
The code size reduces proportionally to the number of functions removed. Furthermore, the number of
unique ROP gadgets is reduced significantly, which narrows down the choices an attacker has when exploiting
a vulnerability. While an application-specific software stack alone does not prevent code-reuse attacks,
the combination of a tailored software stack with other defenses (\eg, CFI) might restrict an attacker
sufficiently to prevent exploitation.

\subsection{Web Applications}
\label{sec:evalwebapps}

\begin{table*}[t]
	\centering
	\caption{Results for PHP. The categories show the number of sensitive functions
		remaining in the PHP interpreter for each configuration. The special configurations \emph{complete} and
		\emph{minimal} give the numbers of sensitive functions for an unmodified PHP interpreter and a
		PHP interpreter containing the least number of functions to be executable.}
	
	\scalebox{1.00}{
		\begin{tabular}{l r r r r r}
			\toprule
			
			& \multicolumn{2}{c}{\tabh{Base Interpreter}}
			& \multicolumn{3}{c}{\tabh{Application-Specific Interpreter}} \\
			
			\cmidrule(lr){2-3}
			\cmidrule(lr){4-6}
			
			{}                      &  Complete &  Minimal &  FluxBB &  OpenConf &  Wordpress \\
			\midrule
			Code Execution          &         5 &        0 &       3 &         2 &          3 \\
			Command Execution       &         7 &        0 &       1 &         1 &          4 \\
			
			\bottomrule
		\end{tabular}
	}
	
	\label{tab:phpvulns}
\end{table*}

To show the applicability of an application-specific software stack for applications using
a script interpreter, we measure the impact of our approach on web applications, namely 
FluxBB (version 1.5.10, 21,295 LOC), OpenConf (version 6.80, 21,232 LOC), and Wordpress (version 4.9.1, 183,820 LOC).
We focus on web applications for PHP and use
the same interpreter as compiled for the evaluation in Section~\ref{sec:evallib}.
To give a realistic overview, we have chosen web applications of different categories and sizes.
To generate the initial whitelist of PHP functions as described in Section~\ref{sec:approachphp},
we use the static analysis tool Parse~\cite{parse}. Unfortunately, Parse does not support
the paradigm of object-oriented programming,
which leads to the necessity to add two additional functions to the initial whitelist for \emph{FluxBB}
(\texttt{dir} and \texttt{read})
and one for \emph{Wordpress} (\texttt{mysqli\_connect}).

Although modern web applications often provide a way to install additional plugins,
we only evaluate our approach on the basic web applications to give a base line of
removable functionalities. If someone wants to use specific plugins,
these plugins only have to be included into the extraction of PHP functions for the initial whitelist
to work with the resulting customized interpreter.

To evaluate the quality of the removed code, we measure the number of remaining sensitive functions
in the script interpreter. We use the categories provided by the open source version of RIPS, a static PHP security scanner~\cite{ripssinks}. 
Since the goal of an application-specific script interpreter is to reduce the impact of an attacker executing arbitrary PHP code
(\eg, by uploading an attacker controlled script file), we focus on the categories \emph{Code Execution} and \emph{Command Execution}.
\emph{Code Execution} contains all functions that allow an attacker to execute arbitrary PHP functionality
and \emph{Command Execution} contains all functions that allow an attacker to execute shell commands on the host.
Table~\ref{tab:phpvulns} shows the full results, in the following we provide a high-level overview.

The base interpreter without any functions removed has five PHP functions in the \emph{Code Execution} category
(\texttt{assert}, \texttt{create\_function}, \texttt{preg\_filter}, \texttt{preg\_replace}, and \texttt{preg\_replace\_callback}).
In contrast, a minimal configuration
of the interpreter (least amount of PHP functions necessary to run the interpreter itself) does not have any such function.
This shows that it is possible to remove this functionality completely from the interpreter as long as the
target web application does not use one of the sensitive functions.
Unfortunately, all projects use some \emph{Code Execution} functionality and hence our approach is not able
to remove it completely from the script interpreter with \emph{FluxBB} using three different PHP functions, \emph{OpenConf} two,
and \emph{Wordpress} three.

PHP functions that provide the ability to execute arbitrary shell commands on the host system are in the category
\emph{Command Execution}. A complete PHP interpreter provides seven such functions
(\texttt{exec}, \texttt{passthru}, \texttt{popen}, \texttt{proc\_open}, \texttt{shell\_exec}, \texttt{system}, and \texttt{mail}) and a minimal configuration none.
Unfortunately, each of the web applications of our evaluation set again uses
at least one sensitive function from the category.
For a \emph{FluxBB} installation, the only PHP function allowing arbitrary shell command execution remaining is \texttt{exec}.
However, since \texttt{exec} is only used to display the system's uptime in the administration control panel,
removing it from the code would allow to remove the ability to execute shell commands completely from the script interpreter.
Hence, an attacker that is able to upload her own script file to a web server is no longer able to execute shell commands.
An \emph{OpenConf} configuration has also only one PHP function remaining in the \emph{Command Execution} category, the function \texttt{mail}.
However, there are multiple limiting factors to consider before an attacker is able to execute shell commands 
with the help of \texttt{mail} which we discuss in Section~\ref{sec:casestudies} in detail.
Hence, a tailored script interpreter for \emph{OpenConf} removes the attack vector of \emph{Command Execution} in most cases completely.
A configuration for \emph{Wordpress} has still four PHP functions that allow shell command execution.
Here, the functionality still remains in the script interpreter and a malicious usage is only mitigated by the insertion of booby traps as explained in Section~\ref{sec:booby}.
An attacker not knowing about the tailored PHP interpreter that gains arbitrary PHP function execution could trigger
a booby trap by executing a removed functionality.

In summary, an application-specific script interpreter reduces the available options for executing code or shell commands.
Furthermore, it is also able to remove certain functionalities altogether and leave the attacker with no possibility to
perform such an attack. In cases where the functionality still remains in the interpreter, it mitigates its malicious effects by
inserting booby traps (which are especially effective in case of proprietary web applications) that can be triggered by an attacker using a removed functionality.

\subsection{Use Case: Wordpress Container}
\label{sec:use_case_wordpress}

\enlargethispage{1cm}
To evaluate the debloating effect for a real-world scenario, we created a Docker container
for our running example, an application-specific \emph{Wordpress} installation. This container comprises
of a PHP interpreter tailored to \emph{Wordpress}, as well as a \emph{musl-libc} tailored to
the \emph{Nginx} web server and PHP interpreter. Since the web server has to interact with
the interpreter directly, PHP is additionally compiled with the FastCGI Process Manager (FPM).
This scenario comprises a setting for which our approach was designed. One shared library tailored to
multiple applications to keep the usability benefits of dynamic linking and a script interpreter customized
for a web application.

The code reduction for the script interpreter is as discussed in Section~\ref{sec:evalwebapps}.
However, the reduction in the library is different since it is now tailored to two applications.
The code of the \emph{musl-libc} is reduced to 351 kB (34.9\% of its original size).
To put things in perspective, the \emph{musl-libc} tailored solely to a \emph{Wordpress} customized
PHP interpreter has only 33.4\% of its code remaining and a \emph{Nginx}-specific library
27.4\%. This suggests that most of the library functions are shared by PHP and \emph{Nginx}.
Only 2.958 unique ROP gadgets were found (41.2\% of the original amount).
Even when comparing to a library specific to a complete PHP interpreter, this shared \emph{musl-libc} setup results in a
smaller library with less code.

In summary, this real-world setting shows a significant code reduction
even with a library tailored to multiple applications. Since this code reduction restricts the options 
for an attacker performing an attack (\eg, whole function reuse, ROP, or PHP code execution),
it is an important additional piece for a security-in-depth environment already providing other forms of
defenses (\eg, CFI).

\subsection{Security Evaluation}
\label{sec:casestudies}

\emph{OpenConf} 5.30 had multiple vulnerabilities that could be chained together to gain remote code execution~\cite{openconfrce}.
This was achieved by injecting PHP code into an uploaded file and executing it.
In an application-specific script interpreter for \emph{OpenConf}, the attacker's possibilities are
limited after gaining PHP code execution. 
The only remaining way to execute shell commands is by using the \texttt{mail} function which allows control over
the arguments passed to the underlying sendmail command.
However, before the arguments are passed to sendmail by the PHP interpreter, they are escaped internally.
As a result, it is exploited by creating a file that can be abused as PHP shell
and thus gain PHP code execution~\cite{phpmailexploit}.
However, again the only remaining way for the attacker to execute shell commands with her created PHP shell is
with the \texttt{mail} function. Hence, it is not possible for the attacker to execute any shell commands
with the tailored PHP interpreter.
The only exception is a system that uses the Exim mail server which allows a direct shell command execution with the \texttt{mail} function. Therefore, depending on the system configuration, an application-specific script interpreter would mitigate such an attack.

CVE-2016-5771 and CVE-2016-5773 in the PHP interpreter were found for Pornhub's bug bounty program in 2016~\cite{php-rce}.
The penetration testers used it to exploit the \texttt{unserialize} function and gain remote code execution on the server.
In their ROP shellcode, they used the function \texttt{zend\_eval\_string} to interpret a given string as PHP code.
Although an application-specific PHP interpreter would not have eliminated this vulnerability (since the code
was used by the web application), the exploiting could be made more difficult with it. For example, the native code function
\texttt{zend\_eval\_string} is not present in any of our tailored interpreter instances (except the complete PHP interpreter).
Additionally, when interpreting a string as PHP code, it might use a removed functionality and thus trigger a booby trap.
Hence, depending on the used web application, the range of suitable candidates to use for an exploit can be limited.

\subsection{Performance}

Since our approach only removes unnecessary code from the support layer of the target application,
it does not induce a performance penalty. However, it does not have a performance gain either, because
only code is removed that is not executed by the application anyways.
The memory consumption of an application-specific library is smaller than the consumption of the complete library,
since code is removed from the binary and therefore not loaded into memory. Nonetheless, since each group of applications
need their own tailored library, the overall
memory consumption of the system is increased. However, since using containers for each service 
(which also increase the memory consumption for each used library) gains more popularity, we deem it acceptable for practical deployments.

\section{Discussion}
\label{sec:discussion}

Scripting languages often offer the possibility to dynamically evaluate code (such as \texttt{eval} in PHP).
When used by the application, it makes the initial analysis
to gather all necessary interpreter functionalities much harder.
Our approach relies on the accuracy of specialized analysis tools for this.
However, if the analysis tool is not able to provide accurate data, the
tailored interpreter could break the application.
Furthermore, if a user-provided input is directly passed to an evaluation
function, stripping down the interpreter becomes impossible since
the user can provide any programming construct she likes.
However, such flawed code constructs allow direct access to the system anyway
and trying to prevent it can be regarded as a losing battle.

As evident from our evaluation, an application-specific interpreter
reduces the options an attacker has if she is able to execute own code in a targeted web application.
Furthermore, it is able to remove certain vulnerability classes completely.
However, if a web application uses a certain interpreter functionality
that can also be used for an attack, our approach is not able to
thwart this. To be more precise, if a web application relies on
the PHP function \texttt{exec} to execute commands directly on the system
(like in the case of \emph{FluxBB}),
our approach cannot remove it. To mitigate attacks using
this functionality, approaches to monitor such remaining functions
can be deployed additionally~\cite{staicu2018synode}.

We showed that the concept of our approach is capable of working with script interpreters
such as PHP and Ruby. However, as script interpreters have different internal structures,
our approach cannot be used directly with another interpreter such as Python. To
support it, domain knowledge of the interpreter's internal workings has to be integrated
(\ie, the mapping of script functions to native code functions).
As this merely means that additional engineering effort is needed to support other interpreters,
it does not constitute a limitation of the general concept of our approach.

Another limitation is that each application needs
its own customized libraries.
As a result, when running multiple services like a web application
in combination with a database server, both need their own tailored libc
(or combine their analysis results to create one libc for both applications).
On first glance, this seems infeasible for a real-world scenario.
However, the recent trend to separate each part of a service into a container,
such as Docker\cite{docker} (which uses Alpine Linux with \emph{musl-libc} for official containers),
makes our approach applicable for real-world scenarios.
When running a web application, one container can contain the web server as
well as a script interpreter (\eg, PHP) with a shared application-specific software stack
and another container the database server with its own tailored software stack.
Thus, enhancing the security mechanism of separating services with reduced options for an attacker
to reuse existing code.

As the evaluation in Section~\ref{sec:eval} has shown, minor manual configuration is still necessary
in some cases. For web applications these were cases where the used static analysis tool Parse
was not able to process object-oriented programming constructs. However, this is not a shortcoming
of our approach, but just a limitation of the used analysis tool. Using a different analysis tool
that is capable of handling object-oriented programming like RIPS~\cite{rips} solves this problem.
Minor manual configuration was also necessary for both tested libc versions. These were either cases
that LLVM could not handle due to assembly,
functions that are called by the loader, or functions
that were resolved dynamically during runtime by the loader as explained in Section~\ref{sec:manualconf}.
These cases require more engineering work and do not constitute conceptual limitations of our approach.
Assembly directly used in the source code can either be lifted to LLVM IR with tools such as McSema~\cite{mcsema}
or processed separately. Entry point functions called directly by the loader can be whitelisted initially by
just adding the names of the C specific starting functions (\eg, \texttt{\_start}). We did not do this to have
a complete evaluation.
Dynamically resolved functions can be addressed by integrating a data-flow analysis which ends in the corresponding
library functions (\eg, \texttt{dlsym}).
However, solving this in general is hard since the only case we encountered used a self-implemented
function of the \texttt{dlsym} functionality to resolve the function pointer.
Hence, our approach can be seen as a first step to an automated way to create application-specific software stacks.

An additional use case for our approach are restricted script interpreter environments that execute user provided untrusted scripts such as Google App Engine~\cite{googleappengine}. These script interpreters prevent internally the usage of specific sensitive functions from being executed. However, Park~et~al.~\cite{park2018bytecode} presented an attack with a restricted attacker model that is able to rewrite bytecode of functions to execute these sensitive functions and therefore bypassing the restriction. By applying our approach for an application-specific script interpreter, these restricted functionalities are completely removed from the interpreter and hence such an attack cannot use them.

Our current prototype focuses on removing unused code from shared libraries and script interpreters written in C,
however, support for C++ is subject of future work. To work with C++, our approach has to be able
to handle virtual function tables (vtables) which are used on a low-level to implement polymorphism.
A naive approach would be to whitelist all functions that are part of a vtable. However, this would
decrease the precision of the code debloating and heavily overestimates the used functions. A better way
would be to improve the static analysis to only keep functions in the vtable that are actually used.
For this to work correctly, our approach has to track the data flow of vtables precisely to identify all used functions
and must be able to modify entries in the vtables to remove unused ones~\cite{pawlowski2017marx}.

Our approach uses a flow-insensitive analysis to find function pointer targets with which we did not
encounter any misses during our evaluation. However, the C programming language allows constructs 
that do not provide sufficient meaningful information in LLVM to determine the possible targets.
In these edge cases, a more sophisticated points-to analysis has to be implemented like the one developed by
Emami~et~al.~\cite{emami1994context}.

\section{Related Work}

Debloating software is an appealing approach to thwart attacks and we now discuss works closely related 
to ours. Based on the observation that an application only uses a small part of the code provided
by a shared library, Quach~et~al.~\cite{quach2018debloating} presented a debloating approach.
They developed a compiler extension that adds metadata to an ELF binary (application and shared libraries)
about the location of functions and their dependencies. On execution of an application, the loader writes
the shared library into memory and then removes all functions that are not used by the application by overwriting them.
However, though the analysis is similar to the presented one, their approach is only applicable to native code applications and does not work with applications written for a script interpreter.

JRed~\cite{jiang2016jred} is an automated approach to remove unused code from Java applications.
It analyzes the bytecode of an application and removes unused code in the application itself and core
libraries of the JRE. However, it is only capable of handling Java bytecode and ignores native code
libraries during its analysis. Since JRed only targets Java bytecode, it does not tackle challenges like indirect
control-flow transfers through function pointers as done by our approach.
Landsborough~et~al.~\cite{landsborough2015removing} presented an approach to remove unwanted
functionalities from binary code by using a genetic algorithm.
Since it works on traces obtained via dynamic analysis, it needs test cases that execute every
functionality the target application should keep. If the set of test cases is not complete,
the code corresponding to a needed but not tested functionality is removed and thus breaks the application.
Additionally, it does not scale and did not even terminate when removing a
feature from the \emph{echo} application of coreutils.
\emph{Chisel}~\cite{heo2018effective} aims to support programmers to debloat programs. It needs the
source code and a high-level specification of its functionalities to remove unwanted features
with the help of delta debugging. A similar goal is pursued by Sharif~et~al.~\cite{sharif2018trimmer} and
their prototype implementation \emph{TRIMMER}, a LLVM compiler extension. With the help of a user-provided manifest
about the desired features, it tries to remove unwanted functionalities to debloat the application.
A binary-only approach targeting specifically applications using a client-server architecture is presented
by Chen~et~al.~\cite{chen2018toss}. Their approach uses binary-rewriting techniques and a user-provided list
of features with corresponding test cases to execute those to customize the target application.
\emph{BinRec}~\cite{kroes2018binrec} also aims at debloating already compiled applications.
It is based on LLVM and needs to lift the target binary into the LLVM IR before it can perform its transformations.
Since automatically removing features from an application on the binary level is prone to errors, \emph{BinRec} also provides a
fallback mechanism to use removed code from the original binary.
In contrast to our approach, these approaches focus on removing features
from a target application itself, while we aim to remove unused functionalities from libraries and script interpreters.

An approach to debloat the Linux kernel was presented by
Kurmus~et~al.~\cite{kurmus2013attack}. Their
approach focuses on optimizing the configuration for the Linux kernel to remove
unnecessary features at compile time. This work is orthogonal to ours and can further
improve the security of the system by not only tailoring the userspace software stack in an application-specific way,
but also optimizing the Linux kernel to target a specific application.

\section{Conclusion}

In this paper, we presented an approach to compile shared libraries
tailored to a specific application by removing unused code from them.
Since complex applications, such as the PHP interpreter, do not
even use half of the provided functions in a shared library,
we showed that this debloating significantly reduces the choices an attacker has for code-reuse attacks.
Furthermore, we demonstrated that with the help of domain
knowledge, our approach is also capable of tailoring a script interpreter
to a script application (\eg, a web application).

We demonstrated an application-specific software stack tailored to a
\emph{Wordpress} installation (customized PHP interpreter, \emph{libc} tailored
to web server and interpreter), and showed a significant code reduction.

\section*{Acknowledgements}

This work was supported by the German Research Foundation (DFG) within the
framework of the Excellence Strategy of the Federal Government and the States
-- EXC~2092 \textsc{CaSa} -- 39078197. In addition, this work was supported by
the European Research Council (ERC) under the European Union’s
Horizon 2020 research and innovation programme (ERC Starting
Grant No. 640110 (BASTION)).

\bibliographystyle{plain}
\bibliography{paper}

\appendix

\subsection{Algorithm}

The complete algorithm of our approach that is capable of handling shared libraries,
PHP and Ruby interpreter is given in Algorithm~\ref{alg:complete}. 
A prototype implementation of this algorithm is available at
\url{https://github.com/RUB-SysSec/ASSS}.

\begin{algorithm*}[t]
	\caption{Complete algorithm to whitelist all needed functions during compilation which is capable of handling
		     shared libraries, the PHP interpreter and the Ruby interpreter.}
	\label{alg:complete}
	\setlength{\columnseprule}{0.4pt}
	\begin{multicols}{2}
		\begin{algorithmic}
			\State global set $whitelist$
			\State global set $php\_whitelist$
			\State global set $ruby\_whitelist$
			\\			
			\\

			\Function{visit\_module}{$module$}
			\State // Start Global Exploration
			\ForAll{$global$ in $module.globals$}
			\State \Call{explore\_global}{$global$}
			\EndFor
			\\
			\State // Start Function Exploration
			\ForAll{$func$ in $module.funcs$}
			\If{$func$ in $whitelist$}
			\State \Call{explore\_function}{$func$}
			\EndIf
			\EndFor
			\EndFunction
			\\
			\\
			\\
			\\
			
			\State \textbf{// Global Exploration}
			\Function{explore\_global}{$global$}
			\If{is php \textbf{and} $global$ is php function table}
			\ForAll{($func\_name$, $func\_ptr$) in $global$}
			\If{$func\_name$ not in $php\_whitelist$}
			\State $global$.delete($func\_name$, $func\_ptr$)
			\Else
			\State $whitelist$.insert($func\_ptr$)
			\EndIf
			\EndFor
			\ElsIf{$global$ is struct}
			\ForAll{$member$ in $global$}
			\State \Call{explore\_global}{$member$}
			\EndFor
			\ElsIf{$global$ is function pointer}
			\State $whitelist$.insert($global$)
			\EndIf
			\EndFunction
			\\
			\\
			\\

			\State \textbf{// Function Exploration}
			\Function{explore\_function}{$func$}
			\State $whitelist$.insert($func$)
			\State set $targets$
			\\
			\ForAll{$instr$ in $func$}
			\If{$instr.type$ == call}
			\State $targets$.insert($instr.target$)
			\If{is ruby}
			\If{ $instr.target$ is ruby register function}
			\State ($func\_name$, $func\_ptr$) = $instr.args$
			\If{$func\_name$ in $ruby\_whitelist$}
			\State $targets$.insert($func\_ptr$)
			\Else
			\State delete instruction $instr$
			\EndIf
			\EndIf
			\EndIf
			
			\ElsIf{$instr.type$ == store}
			\State $targets$.insert($instr.store\_value$)
			\ElsIf{$instr.type$ == return}
			\State $targets$.insert($instr.return\_value$)
			\ElsIf{$instr.type$ == select}
			\State $targets$.insert($instr.true\_value$)
			\State $targets$.insert($instr.false\_value$)
			\ElsIf{$instr.type$ == phi}
			\ForAll{$value$ in $instr.incoming\_values$}
			\State $target$.insert($value$)
			\EndFor
			\EndIf
			\EndFor
			\\
			\ForAll{$target$ in $targets$}
			\If{$target$ is function}
			\If{$target$ not in $whitelist$}
			\State \Call{explore\_function}{$target$}
			\EndIf
			\EndIf
			\EndFor
			\EndFunction
		\end{algorithmic}
	\end{multicols}
\end{algorithm*}

\end{document}